\documentclass[aps,pre,
reprint,
notitlepage,
showpacs,floatfix,nofootinbib,
superscriptaddress
]{revtex4-2}
 \usepackage{subfigure}
 \usepackage{amssymb}
 \usepackage{amsfonts}
 \usepackage{amsmath}
 \usepackage{amsthm}
 \usepackage{epsfig}
 \usepackage{float}
 \usepackage[usenames,dvipsnames]{color}

 \usepackage{graphics,graphicx,xcolor,subfigure}

 \usepackage[hidelinks,unicode=true]{hyperref}
 \hypersetup{colorlinks=true,
 	linkcolor=blue,
 	urlcolor=blue,
 	citecolor=blue,
 	pdfhighlight=/N
 }
 
 \usepackage{comment}
 \usepackage[normalem]{ulem}

\newcommand{\av}[1]{\langle {#1} \rangle}

\newcommand{\Ninf}{N_\mathrm{inf}}

\newcommand{\NSI}{N_\mathrm{SI}}

\begin{document}

\title{Comparison of theoretical approaches for epidemic processes with waning immunity in complex networks}
\author{Jos\'e Carlos M. Silva}
\affiliation{Departamento de F\'{\i}sica, Universidade Federal de Vi\c{c}osa, 36570-900 Vi\c{c}osa, Minas Gerais, Brazil}

\author{Diogo H. Silva}
\affiliation{Instituto de Ci\^{e}ncias Matem\'{a}ticas e de Computa\c{c}ão, Universidade de S\~{a}o Paulo, S\~{a}o Carlos, SP 13566-590, Brazil}

\author{Francisco A. Rodrigues}
\affiliation{Instituto de Ci\^{e}ncias Matem\'{a}ticas e de Computa\c{c}ão, Universidade de S\~{a}o Paulo, S\~{a}o Carlos, SP 13566-590, Brazil}

\author{Silvio C. Ferreira}
\affiliation{Departamento de F\'{\i}sica, Universidade Federal de Vi\c{c}osa, 36570-900 Vi\c{c}osa, Minas Gerais, Brazil}
\affiliation{National Institute of Science and Technology for Complex Systems, 22290-180, Rio de Janeiro, Brazil}

\begin{abstract} 
The role of waning immunity in basic epidemic models on networks has been undervalued while being noticeable fundamental for real epidemic outbreaks. One central question is which mean-field approach is more accurate in describing the epidemic dynamics. We tackled this problem considering the susceptible-infected-recovered-susceptible (SIRS) epidemic model on networks. Two pairwise mean-field theories, one based on recurrent dynamical message-passing (rDMP) and the other on the pair quenched mean-field theory (PQMF), are compared with extensive stochastic simulations on large networks of different levels of heterogeneity. For waning immunity times longer than or comparable with the recovering time, rDMP outperforms PQMF theory on power-law networks with degree distribution $P(k)\sim k^{-\gamma}$. In particular, for  $\gamma>3$, the epidemic threshold observed in simulations is finite, in qualitative agreement with rDMP, while PQMF leads to an asymptotically null threshold. The critical epidemic prevalence for $\gamma>3$  is localized in a finite set of vertices in the case of the PQMF theory. In contrast, the localization happens in a subextensive fraction of the network in rDMP theory. Simulations, however, indicate that localization patterns of the actual epidemic lay between the two mean-field theories, and improved theoretical approaches are necessary to understanding the SIRS dynamics.

\end{abstract}

\maketitle

\section{Introduction}

Several epidemic outbreaks threaten humanity by spreading throughout the globe, such as avian influenza in Southeast Asia and Western Europe~\cite{Colizza2007}, Ebola in West Africa~\cite{Gomes2014},  Zika virus in the Americas~\cite{Angel2017} and, more recently, the ongoing COVID-19 pandemic~\cite{Desai2019,Arenas2020,Guilherme2020}. Understanding the immune response is essential to  identify vulnerable groups~\cite{Levin2020,Verity2020,Cerqueira_Silva2021}, develop vaccines with high efficacy, and construct optimal distribution strategies~\cite{schulenburg2022}. 
The role of waning immunity in epidemic models is essential to understand the long-term evolution of an infectious disease. Besides shedding light on the dynamics of spreading pathogens, the development of accurate theoretical frameworks may lead to improved forecasting. Perfect immunity response is assumed in the susceptible-infected-removed (SIR) model~\cite{Anderson2000}, in which susceptible individuals are infected with rate $\lambda$ upon each contact with a contagious individual and heal spontaneously with the rate $\mu$ remaining in a recovered state where she or he cannot be reinfected. Conversely, if no immunity is conferred and an infected individual becomes susceptible again immediately after healing, the susceptible-infected-susceptible (SIS)~\cite{Anderson2000} epidemic model is essential. In the case of waning immunity with an average time $1/\alpha$ after recovering, the susceptible-infected-recovered-susceptible (SIRS)~\cite{Anderson2000}  dynamics is the fundamental process.

From a theoretical perspective, dynamical message-passing (DMP) theory suits very well the SIR model~\cite{Karrer2010}  where the transmission events are described by ``messages'' that do not backtrack in consonance with the permanent immunity. When applied to dynamics on top of tree-like networks, it gives an exact description, while DMP theory yields upper bounds to the outbreak sizes in non-tree-like networks. The DMP theory, such as other theoretical frameworks, allows us to describe the importance of spreaders in the underlying dynamics~\cite{Castellano2017,Torres2021}.

The nature of the activation process in the SIS dynamics on random power-law networks involves a feedback mechanism where the epidemic activity within subextensive subgraphs is self-sustained and spreads the activity to the rest of the network~\cite{Boguna2013,Sander2016,Castellano2012}. The quenched mean-field theory (QMF)~\cite{Wang2003, Chakrabarti2008, VanMieghen2009, VanMieghem2012_a, VanMieghem2012_b, Romualdo2015}, in which the full network structure is explicitly considered, describes qualitatively the vanishing of the epidemic threshold in random networks presenting power-law degree distribution, $P(k)\sim k ^{-\gamma}$, regardless the value of $\gamma$~\cite{Wang2003,Chakrabarti2008}. The accuracy of such predictions can be improved by adding dynamical correlation in a pairwise level~\cite{Mata2013,Diogo2019}, the pair quenched mean-field (PQMF) theory, especially in the regime of high epidemic prevalence~\cite{Diogo2020}. Shrestha et al. \cite{Munik2015} proposed a modified version of the  DMP theory for dynamic processes with an active (fluctuating) steady state in the now called recurrent dynamical message passing (rDMP) theory. However, Castellano and Pastor-Satorras~\cite{Claudio2018} argued that, by construction, the backtracking mechanism absent in rDMP theory is essential to the activation of the SIS dynamics in heterogeneous networks~\cite{Mountford2013,Chatterjee2009,Boguna2013,Sander2016}. So, rDMP theory is not a suitable approach and notably worser for degree exponents $\gamma>2.5$ when the dynamics is ruled by self-sustained activation of hubs~\cite{Boguna2013}. 

The  SIS and SIR dynamics are limit cases of the SIRS model when $\alpha \rightarrow \infty$ and $\alpha \rightarrow 0$, respectively. The SIRS and SIS dynamics share the same symmetries on lattices and belong to the directed percolation universality class~\cite{DeSouza2010,Joo2004}. In power-law networks, standard mean-field theories predict the same epidemic threshold and critical behavior as the SIS~\cite{Ferreira2016,Bancal2010}. However,  the waning immunity is capable of modifying the epidemic activity in isolated hubs, implying that the activation mechanism of the SIS model, based on long-range mutual infection of hubs, is depleted~\cite{Ferreira2016} and the mechanism is altered. The epidemic threshold, above which an outbreak can reach a finite fraction of the population in the thermodynamic limit, is a fundamental epidemiological parameter. In random scale-free networks with degree exponent $\gamma<5/2$ the three aforementioned models (SIR, SIS, and SIRS) behave very similarly with the epidemic threshold very well described by the QMF theory~\cite{Ferreira2016}. However, remarkable differences emerge for $\gamma>5/2$ and especially for $\gamma>3$. For example, the asymptotic epidemic threshold of the SIS models is null for any value of $\gamma$~\cite{Boguna2013,Chatterjee2009,Ferreira2012}, while the threshold is finite for $\gamma>3$ and a phase transition is observed  in SIR and SIRS with finite $\alpha$~\cite{Ferreira2016}.

The QMF theory and its improved PQMF version for SIS dynamics also deviate from simulations very near to the epidemic threshold~\cite{Diogo2019,Diogo2020}. The accuracy of these theories is related to spectral localization of the Jacobian matrices obtained in stability analysis of the absorbing state~\cite{Castellano2017,Diogo2019,Goltsev2012}. However, PQMF theory for SIS has shown to be very accurate if the analysis is not too close to the epidemic threshold~\cite{Diogo2020}.

Despite its natural relevance for applications, the SIRS dynamics has attracted much less attention than its SIR and SIS limits, and an efficient theoretical approximation for SIRS dynamics on networks remains an open question. In the present work, the role of immunity is investigated using intermediate values of the waning immunity rate $\alpha$. We compared extensive stochastic simulations with rDMP and PQMF  theories, establishing which theory performs better. In the case of power-law networks, the rDMP theory correctly predicts the epidemic threshold behavior (vanishing or not) at the limit of asymptotically large networks, while the PQMF theory is ruled by localization on a finite set of vertices which leads to a vanishing threshold for $\gamma>3$, in contrast with stochastic simulations that indicate a finite threshold. However, we also report evidence that the rDMP theory underestimates while PQMF overestimates the epidemic localization leading, respectively, to upper and lower bounds for the epidemic thresholds of the actual SIRS dynamics. Our results indicate that an improved theoretical approach is necessary to accurately describe the critical behavior of the SIRS dynamics on networks.   

The remainder of this paper is organized as follows. In Sec.~\ref{sec:theory}, the theoretical approaches for the SIRS model are presented. The epidemic thresholds obtained with stochastic simulations are compared with the theoretical predictions for different complex networks in Sec.~\ref{sec:result}. Finally, in Sec.~\ref{sec:conclusions}, we present our conclusions and prospects. Two appendices with technical details of the work complement the paper.

\section{Theoretical approaches for the SIRS model on networks}
\label{sec:theory}

\subsection{The PQMF theory}
\label{sec:mean_field}

In the PQMF theory, the whole network structure is considered and the dynamical correlations are partially taken into account in a pairwise level. We extend the analysis performed to the SIS dynamics in Ref.~\cite{Mata2013} to the SIRS model. The state of every node $i$ is presented by $\sigma_i=0$ (susceptible), 1 (infected), or 2 (recovered). We define the variables $s_{i}=[0_{i}]$, $\rho_{i}=[1_{i}]$, $r_{i}=[2_{i}]$, representing the marginal probabilities of finding a node $i$ in the susceptible, infected, and recovered states, respectively. We also represent the probabilities that a pair $(i,j)$ assumes states $[\sigma_i,\sigma_j]$ by   $\theta_{ij}=[2_{i},1_{j}]$, $\chi_{ij}=[2_{i},0_{j}]$, $\phi_{ij}=[0_{i},1_{j}]$,   
$\psi_{ij}=[1_{i},1_{j}]$, 
$\omega_{ij}=[0_{i},0_{j}]$,
$\upsilon_{ij}=[2_{i},2_{j}]$, 
$\overline{\theta}_{ij}=[1_{i},2_{j}]$, $\overline{\chi}_{ij}=[0_{i},2_{j}]$, and  $\overline{\phi}_{ij}=[1_{i},0_{j}]$. They represent the joint probabilities of finding two neighbor nodes $i$ and $j$ in combinations of states allowed by the model. The following closure relations hold for any pair of nodes 
\begin{eqnarray}
s_i =\omega_{ij}+\overline{\chi}_{ij}+\phi_{ij},\nonumber \\ 
\rho_i =\psi_{ij}+\overline{\phi}_{ij}+\overline{\theta}_{ij}, \nonumber \\ 
r_i = \upsilon_{ij}+\theta_{ij}+\chi_{ij}. 
\label{eq:PQMF1}
\end{eqnarray}

The set of equations describing the temporal evolution of the infected and recovered populations are given by
\begin{equation}
	\frac{d\rho_i}{dt} = -\mu \rho_i+\lambda\sum_j A_{ij}\phi_{ij},\,
	\label{eq:PQMF2a}
\end{equation}
and 
\begin{equation}
	\frac{dr_i}{dt} = -\alpha r_i + \mu \rho_i, 
	\label{eq:PQMF2b}
\end{equation}
respectively,  where the adjacency matrix given by $A_{ij}=1$ if $i$ and $j$ are connected and $A_{ij}=0$, otherwise, while the susceptible population is given by $s_i =1-\rho_i - r_i$, in which a constant total population is assumed.  If pairwise dynamical correlations are disregarded, we approximate   $\phi_{ij} \approx s_{i}\rho_{i}$ to obtain the QMF equations
\begin{equation}
\frac{d\rho_i}{dt} = -\mu \rho_i+\lambda s_i\sum_j A_{ij}\rho_{j}.
\label{eq:QMFa}
\end{equation}
As in the HMF theory~\cite{Bancal2010,Ferreira2016}, the epidemic threshold of the SIRS model is the same of the SIS dynamics (see appendix~\ref{app:critical_qmf}), given by the inverse of the largest eigenvalue (LEV) $\Lambda^{(1)}$  of the adjacency matrix~\cite{Chakrabarti2008} associated to its principal eigenvector (PEV)~\cite{Castellano2017,Goltsev2012}:
\begin{equation}
\lambda_\text{c}=\frac{\mu}{\Lambda^{(1)}}.
\end{equation}
Consequently, in random networks presenting power-law degree distributions, this epidemic threshold is  null in the thermodynamic limit when the LEV diverges~\cite{Chakrabarti2008,Wang2003}. As shown in Appendix~\ref{app:critical_qmf}, the fraction of infected nodes  near to the epidemic threshold in SIRS dynamics is proportional to the SIS limit and given by
\begin{equation}
	\rho^{\text{SIRS}}=\left(\frac{\alpha}{\alpha+\mu}\right)\rho^\text{SIS},
	\label{eq:QMF}
\end{equation}    
implying that $\rho\sim (\lambda-\lambda_\text{c})^\beta$ where $\beta^{\text{SIRS}} = \beta^\text{SIS} \equiv 1$~\cite{VanMieghen2009,Goltsev2012}. 

Back to Eqs.~\eqref{eq:PQMF2a} and \eqref{eq:PQMF2b}, the evolution of $\phi_{ij}$   for connected nodes $(A_{ij}=1)$ is given by
\begin{eqnarray} 
\frac{d\phi_{ij}}{dt} & = & -(\mu+\lambda)\phi_{ij}+\alpha\theta_{ij}+
\lambda \sum_{l\neq i}[0_{i},0_{j},1_{l}]A_{lj} \nonumber  \\
& &-\lambda \sum_{l\ne j}[1_{l},0_{i},1_{j}]A_{li}. \label{eq:PQMF3f}
\end{eqnarray}
The interpretation of each term is straightforward. The first term includes both the infection of node $i$ by  $j$ and the spontaneous healing of $j$. The second term is due to the spontaneous waning of immunity of node $i$. The last two terms reckons the infection due to reaming neighbors of $j$ and $i$, respectively. The remaining pairwise equations can be computed as 
\begin{eqnarray} 
\frac{d\theta_{ij}}{dt} & = & \mu\psi_{ij}-(\alpha+\mu)\theta_{ij}+\lambda \sum_{\substack{l\neq i}}[2_{i},0_{j},1_{l}]A_{lj} \label{eq:PQMF3c},\\
\frac{d\chi_{ij}}{dt} & = & \mu\overline{\phi}_{ij}-\alpha\chi_{ij}+\alpha\upsilon_{ij}  -\lambda \sum_{l\neq i}[2_{i},0_{j},1_{l}]A_{lj}, \label{eq:PQMF3d}
\end{eqnarray}
while dynamical equations $\overline{\phi}_{ij}={\phi}_{ji}$, $\overline{\chi}_{ij}={\chi}_{ji}$, and $\overline{\theta}_{ij}={\theta}_{ji}$ can be obtained by switching  $i$ and $j$ in Eqs.~\eqref{eq:PQMF3f}, \eqref{eq:PQMF3c}, and \eqref{eq:PQMF3d}.  Finally, the remaining pair variables $\omega_{ij}$, $\psi_{ij}$ and $\upsilon_{ij}$ can be obtained using relations given in Eq.~\eqref{eq:PQMF1}. To produce a closed system, we approximate the triplets using a pair-approximation~\cite{Avraham1992} 
\begin{equation}
[A_{i}B_{j}C_{k}] \approx \frac{[A_{i}B_{j}][B_{j}C_{k}]}{[B_{j}]}.
\label{eq:PQMF4}
\end{equation}
The closed set of pairwise equations is obtained with Eqs.~\eqref{eq:PQMF1}, \eqref{eq:PQMF2a}, and \eqref{eq:PQMF2b}  joined to
\begin{eqnarray} 
\frac{d\phi_{ij}}{dt} & = & -(\mu+\lambda)\phi_{ij}+\alpha\theta_{ij}+
\lambda \sum_{l\neq i}\frac{\omega_{ij}\phi_{jl}}{s_j} A_{lj} \nonumber  \\
& &-\lambda \sum_{l\ne j}\frac{\phi_{ij}\phi_{il}}{s_i}A_{li}, \label{eq:PQMF4a}
\end{eqnarray}
\begin{equation} 
\frac{d\theta_{ij}}{dt} = \mu\psi_{ij}-(\alpha+\mu)\theta_{ij}+\lambda \sum_{l\neq i}\frac{\chi_{ij} \phi_{jl}}{s_j}A_{lj}, \label{eq:PQMF4b}
\end{equation}
and
\begin{equation} 
\frac{d\chi_{ij}}{dt} = \mu\overline{\phi}_{ij}-\alpha\chi_{ij}+\alpha\upsilon_{ij}  -\lambda \sum_{l\neq i}\frac{\chi_{ij} \phi_{jl}}{s_j}. \label{eq:PQMF4c}
\end{equation}

We can assume that $\rho_{i}\ll 1$ for long times and near to the epidemic threshold,  and so do the other variables which depend on infected or recovered nodes ($r_i$, $\psi_{ij}$, $\phi_{ij}$, \ldots), and consequently $s_i\approx \omega_{ij}\approx 1$. After algebraic handling  to leading order, we obtain the following relation valid for the steady state:
\begin{equation}
\phi_{ij}=\Upsilon \rho_{j}-\Xi\rho_{i},
\label{eq:PQMF5a}
\end{equation}
in which
\begin{eqnarray}
\Upsilon(\mu,\lambda,\alpha)=\frac{2\mu(\mu+\lambda+\alpha)+\lambda\alpha}{2\lambda(\mu+\alpha)+2\mu(\mu+\lambda+\alpha)}
\label{eq:PQMF5b}	
\end{eqnarray}
and
\begin{eqnarray}
\Xi(\mu,\lambda,\alpha)=\frac{\lambda(\alpha+2\mu)}{2\lambda(\alpha+\mu)+ 2\mu(\mu+\lambda+\alpha)}.
\label{eq:PQMF5c} 
\end{eqnarray}

We can now assume a quasi-static approximation where Eq.\eqref{eq:PQMF5a} is plugged in Eq.~\eqref{eq:PQMF2a}, to produce a linear equation with  the Jacobian matrix given by
\begin{eqnarray}
	L_{ij}=-\left[\mu+\lambda k_{i} \Xi(\mu,\lambda,\alpha)\right]\delta_{ij}+\lambda\Upsilon(\mu,\lambda,\alpha) A_{ij},
\label{eq:PQMF6}
\end{eqnarray}
where $\delta_{ij}$ is the Kronecker delta. Thus, using standard linear stability analysis, the absorbing state loses stability and an active steady state emerges when the largest eigenvalue of $L_{ij}$ is null. Equation~\eqref{eq:PQMF6} converges to the SIS Jacobian obtained in Ref.~\cite{Mata2013} when $\alpha \rightarrow \infty$. 

Before analyzing the PQMF theory on general networks, we consider two particular cases: star graph and random regular (RR) networks, 

\subsubsection{Particular case: Star graph}
\label{sec:star_graph}

A star graph is defined as a center $i=0$ connected to $K$ neighbors, $i=1,2,3...K$, of degree $k=1$, represented by the  adjacency $A_{0j}=A_{i0}=1$ and $A_{ij}=0$ otherwise. The  eigenvalues of the Jacobian matrix given by Eq.~\eqref{eq:PQMF6} for the star graph can be computed directly using $\sum_j L_{ij} v_j = \Lambda v_i$ such that the LEV becomes null, providing the epidemic threshold, when
\begin{equation}
	\left(\frac{\lambda_\text{c}}{\mu}\right)^{2}N[\Xi_\text{c}^{2}-\Upsilon_\text{c}^{2}]+\frac{\lambda_\text{c}}{\mu} \Xi_\text{c}(K+1)+1=0,
	\label{eq:pqmf_star2a}
\end{equation}   
where $\Xi_\text{c}$ and $\Upsilon_\text{c}$ are given by Eqs~\eqref{eq:PQMF5b} and~\eqref{eq:PQMF5c} with $\lambda=\lambda_\text{c}$, which are inserted in  Eq.~\eqref{eq:pqmf_star2a} to obtain the epidemic threshold for $K\gg 1$
\begin{equation}
\frac{\lambda_\text{c}}{\mu} \simeq \sqrt{\frac{2\left(\alpha+\mu\right)}{\alpha K}}
\end{equation}
When $\alpha \gg \mu$, the SIS limit for the PQMF theory is recovered~\cite{Mata2013}: ${\lambda_\text{c}}/{\mu} \simeq \sqrt{{2}/{K}}$. On the other hand, when $\alpha\ll \mu$, the epidemic threshold becomes
\begin{equation}
\frac{\lambda_\text{c}}{\mu} \simeq \sqrt{\frac{2\mu}{\alpha K}}.
\end{equation}
The PQMF theory predicts a vanishing epidemic threshold for any nonzero value of the waning immunity rate $\alpha$. This result is in odds with stochastic theory presented in Ref.~\cite{Ferreira2016} and recent rigorous probabilistic analysis of Ref.~\cite{Friedrich2022} where the epidemic lifespan increases algebraically with the graph size $\tau_{K}\sim K^{\alpha/\mu}$ instead of exponentially.

\subsubsection{Particular case: RR network}

In RR networks, all nodes present the same degree $m$, $P(k)=\delta_{k,m}$, and the connections are random. It is direct to check that $v_i=1$ is an eigenvector of Eq.~\eqref{eq:PQMF6} with  eigenvalue 
\begin{equation}
\Lambda= (\mu+\lambda m \Xi)+\lambda m\Upsilon.
\label{eq:pqmf_RRN1}
\end{equation}
Since $v_i>0$ and $A_{ij}$ is positive definite, application of the Perron-Frobenius theorem ensures that it corresponds to the PEV $v_i^{(1)}$. So, the epidemic threshold is obtained when $\Lambda^{(1)}=\Lambda=0$:
\begin{equation}
	\lambda_\text{c}=\frac{\mu (\mu +\alpha)}{(m-1)(\mu+\alpha)-\mu},
	\label{eq:pqmf_RRN2}
\end{equation} 
which is the the same result obtained with the homogeneous pair approximation for SIRS dynamics ~\cite{Joo2004}. The epidemic thresholds of SIS and SIR,  $\lambda_\text{c}^\text{SIS}=\mu/(m-1)$~\cite{Mata2013} and  $\lambda_\text{c}^\text{SIR}=\mu/(m-2)$~\cite{Newman2002}, are obtained in the limits  $\alpha\gg \mu$ and $\alpha\ll \mu$, respectively.

\subsection{The rDMP theory}

In the rDMP approach~\cite{Munik2015}, an infectious node that was infected by a given neighbor  can not reinfect him or herself, which can be encoded by a \textit{message} variable $\rho_{j\rightarrow i}$ defined as  the probability that an infectious node $j$ was infected by any of its  neighbors except node $i$, such that node $j$ can infect $i$, but cannot infect the node which transmitted the contagion to $j$. Assuming a mean-field hypothesis that neglects nearest-neighbor dynamical correlations, this variable evolves as~\cite{Munik2015}    
\begin{equation}
	\frac{d\rho_{j\rightarrow i}}{dt} = -\mu \rho_{j\rightarrow i}+\lambda s_j\sum_{k\neq i} \rho_{k\rightarrow j}A_{jk}.
	\label{eq:rDMP1}
\end{equation}
The remaining compartments of the SIRS dynamics are given by~\cite{Munik2015}
\begin{equation}
\frac{d\rho_i}{dt} = -\mu \rho_i+\lambda s_i\sum_j\rho_{j\rightarrow i},
\label{eq:rDMP2a}
\end{equation}
\begin{equation}
\frac{dr_i}{dt} = -\alpha r_i + \mu \rho_i, 
\label{eq:rDMP2c}
\end{equation}
and $s_i+r_i+\rho_i=1$. Performing again a quasi-static approximation with $\frac{dr_i}{dt}\approx 0$ and linearizing the rDMP equations  around the absorbing state $\rho^{*}_{j\rightarrow i}=0=\rho_j^*$, we obtain  the Jacobian matrix 
\begin{equation}
	J_{j\rightarrow i, k\rightarrow l}=-\mu\delta_{kj}\delta_{ij'}+\lambda B_{j\rightarrow i, k\rightarrow l}
	\label{eq:rDMP3}
\end{equation}
in which,
\begin{equation}
	B_{j\rightarrow i, k\rightarrow l}=\delta_{jl}(1-\delta_{ik})
\end{equation}
is the non-backtracking or Hashimoto matrix~\cite{Florent2013,Hashimoto1989}. The epidemic threshold is then given by the inverse of the LEV of THE Hashimoto matrix~\cite{Munik2015}. 
 
\subsubsection{Particular case: Star graph}

Exploiting the definition of message, $\rho_{j\rightarrow i}$, if $j>0$ is a leaf, it would be infected only by the center such that $\rho_{j\rightarrow 0}=0$. Thus, Eq.~\eqref{eq:rDMP1} with $j=0$ becomes
\begin{equation}
	\frac{d\rho_{0\rightarrow i}}{dt}=-\mu \rho_{0\rightarrow i}+\lambda s_{0} \sum_{k=i}^{K} \rho_{k\rightarrow 0}=-\mu \rho_{0\rightarrow i},
	\label{eq:rDMP_star}
\end{equation}
implying in exponential decay with time, independently of infection rate and that rDMP does not predict an active phase for the SIRS model in a star graph as well as it does not for the SIS dynamics, in odds with both the algebraic and exponential increases of SIRS~\cite{Ferreira2016,Friedrich2022} and   SIS~\cite{Boguna2013,Chatterjee2009} models, respectively. 

\subsubsection{Particular case: RR networks}

For homogeneous networks we have that $\rho_{j\rightarrow i}=\eta$ and Eq.~\eqref{eq:rDMP1} becomes
\begin{equation}
	\frac{d\eta}{dt}=-\mu \eta+\lambda s (m-1)\eta,
	\label{eq:rDMP_RRN1}
\end{equation}
that provides the epidemic threshold  
\begin{equation}
	\lambda_\text{c}=\frac{\mu}{m-1}  
	\label{eq:rDMP_RRN2}	
\end{equation}
and corresponds to the same result of the SIS model and is independent of the rate of waning immunity. One can show that
\begin{equation}
 \rho^\text{SIRS}=\left(\frac{\alpha}{\alpha+\mu}\right)\rho^\text{SIS} = \left(\frac{\alpha}{\alpha+\mu}\right)\frac{\lambda-\lambda_{c}}{\lambda},
\end{equation}
in agreement with Eq.~\eqref{eq:QMF} derived for the  QMF theory.

\section{Theory versus simulation}
\label{sec:result}
To compare the performance of theoretical frameworks, we performed stochastic simulations of the SIRS model following the Gillespie algorithm described in Appendix~\ref{app:algoritmo}. We deal with absorbing state using the quasi-stationary method~\cite{Costa2021,Oliveira2005} explained  in Appendix~\ref{app:algoritmo}. The epidemic threshold is determined using the infection rate corresponding to the largest value of  the dynamical susceptibility defined as $\chi=N(\langle\rho^{2}\rangle-\langle\rho\rangle^{2})/\langle\rho\rangle$~\cite{Ferreira2012}, where the averages are computed in the quasi-stationary regime. We investigate the rate of waning immunity $\alpha\leq\mu$, which corresponds to a time of conferred immunity longer than that of infectiousness. 

Considering star graphs, we have shown that PQMF theory predicts an active state while rDMP does not. Indeed, a stochastic approximation and numerical simulations for SIRS dynamics  on a  star  of size $K+1$  leads to an algebraic increase of the epidemic lifespan given by $\tau_{K}\sim K^{\alpha/\mu}$~\cite{Sander2016} in contrast with the exponential law $\tau_{K}\sim \exp(-\lambda^2K/\mu^2)$ of SIS the dynamics~\cite{Boguna2013}. However, this metastable  activity in star graphs can introduce localization effects on hubs within the networks. In mean-field theories, localization can be investigates computing the inverse participation ratio (IPR) of the PEV, associated to the Jacobian of the corresponding mean-field theory, as a function of network size: adjacency matrix $A_{ij}$ for QMF~\cite{Goltsev2012}, Hashimoto matrix $B_{j\rightarrow,k\rightarrow l}$  for rDMP~\cite{Pastor_Satorras2020}, and Eq.~\eqref{eq:PQMF6} for PQMF~\cite{Diogo2019}. If $\{v_j^{(1)}\}$ are the $N_\text{c}$ components of the normalized PEV, the IPR is defined as~\cite{Goltsev2012}
\begin{equation}
Y_4=\frac{1}{N}\sum_{j=1}^{N_\text{c}} \left[v_j^{(1)}\right]^2.
\label{eq:IPR}
\end{equation}
The IPR scale as $Y_4\sim  N_\text{c}^{-1}$ for a delocalized PEV and becomes a constant larger than zero for localization on a finite subset~\cite{Goltsev2012}. Some systems can present intermediary localization in a subextensive part of the PEV such that $Y_4\sim N_\text{c}^{-\nu}$ with $0<\nu<1$~\cite{Satorras2016}. The localization analysis can be extended to stochastic simulations  by means of the normalized activity vector (NAV) $\{x_i\}$~\cite{Diogo2021} defined in terms of the probability that an node is active (infected) $\rho_i$ as
\begin{equation}
x_i=\frac{\rho_i}{\sqrt{\sum_{j=1}^{N} \rho_j^2}},
\label{eq:NAV}
\end{equation}
and the corresponding IPR can be calculated replacing the PEV by the NAV components in Eq.~\eqref{eq:IPR}.

\subsection{RR networks without and with an outlier}

\begin{figure}[hbt]
	\includegraphics[width=0.8\linewidth]{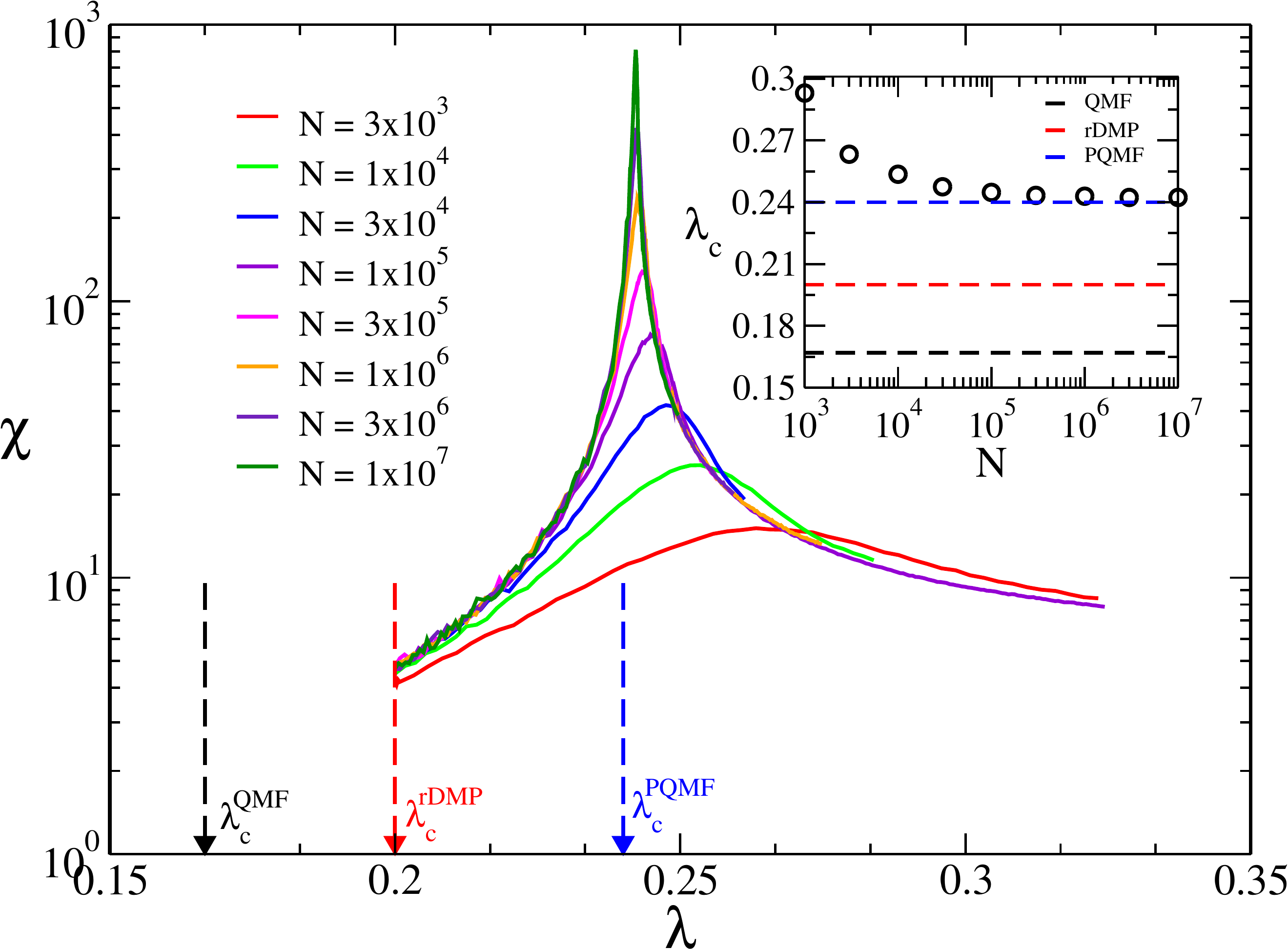}
	\caption{Susceptibility as a function of $\lambda$ for RR networks with $m=6$ and different sizes indicated in the legend. Black, red, and blue dashed arrows correspond to QMF, rDMP and PQMF predictions of the epidemic threshold, respectively. Inset compares the epidemic threshold as a function of the network size in simulations (symbols) and mean-field theories (dashed lines)}
	\label{fig:RRN}
\end{figure}

We start with RR networks since, in principle, these networks present no relevant localization effects. In the thermodynamic limit, the epidemic threshold converges to a finite value as predicted by all theoretical frameworks, for all values of $\alpha$ studied. The PQMF theory outperforms significantly both QMF and rDMP in determining the epidemic threshold in comparison with stochastic simulations, as shown in Fig.~\ref{fig:RRN} for $\alpha/\mu=0.2$. This result holds for other values of $\alpha/\mu$. The partial reckoning of dynamical correlation in rDMP leads to an improvement of the QMF theory, but still substantially below the accuracy of the PQMF theory. The results shown in Fig.~\ref{fig:RRN} correspond to a degree $m=6$. The relative accuracy of all theories is  reduced for lower degrees, and the relative improvement of the PQMF with respect to the other theories is increased.

\begin{figure}[hbt!]
	\includegraphics[width=\linewidth]{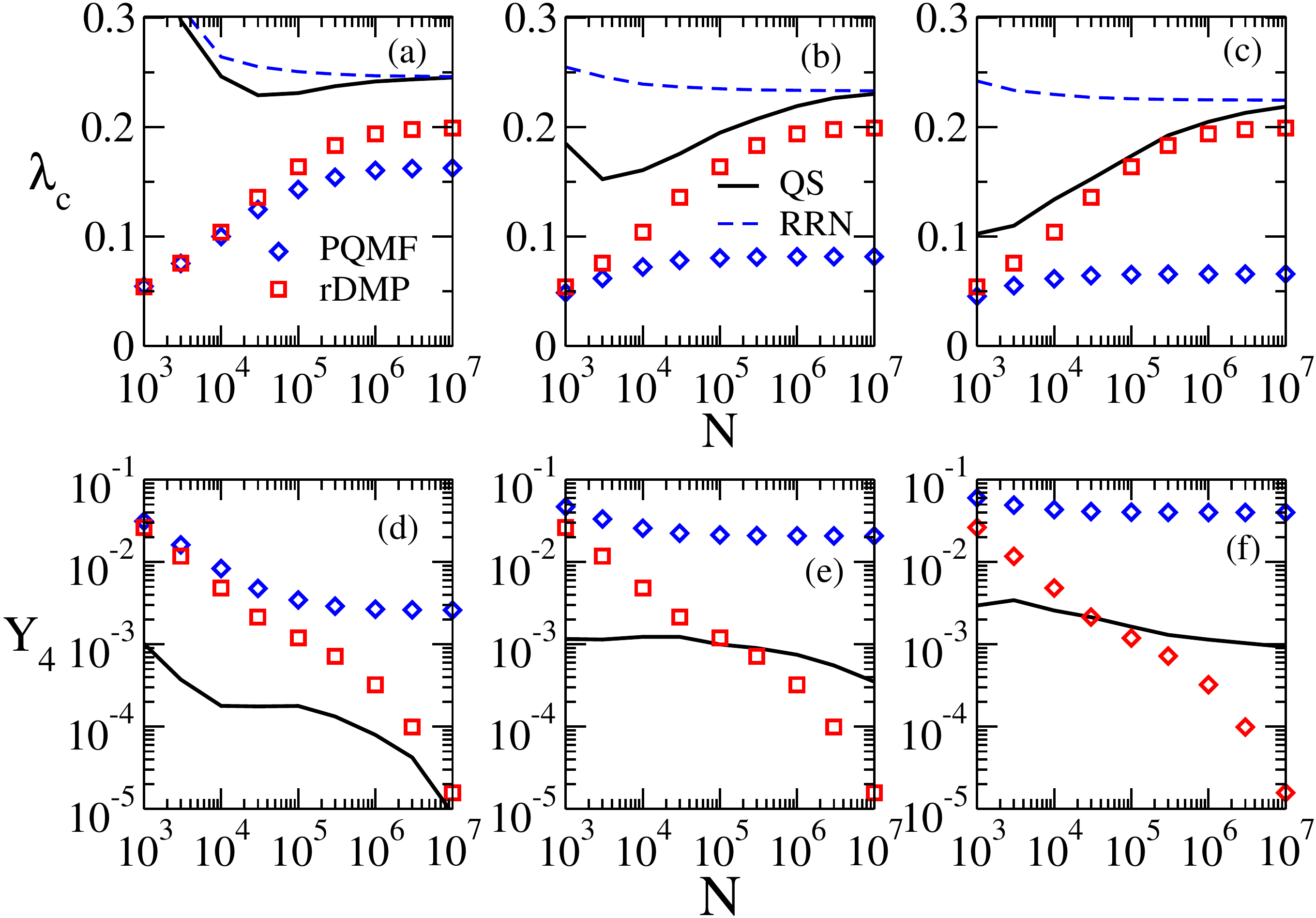}
	\caption{(a,b,c) Epidemic threshold and (d,e,f) IPR as functions of the network size for the mean-field theories and simulations of the SIRS model with (a,d)~$\alpha/\mu=0.1$, (b,e)~$0.5$ and (c,f)~$1.0$. The system is a RR network with $m=6$ plus a single vertex of fixed size $k=10^{3}$. The threshold for a pure RR network is also shown for sake of comparison.}
	\label{fig:hub+RRN}
\end{figure}

We tackle the effects of localization by introducing a single hub with fixed degree $k=10^{3}$ in  a RR network where all the remaining $N-1$ nodes have degree $m=6$. In the SIS model two activation processes, expressed as a double peak at susceptibility curves~\cite{Ferreira2012},  take place: one at the subgraph composed of the hub plus its nearest-neighbors and another at the rest of network coinciding at the epidemic threshold of the pure RR network~\cite{Diogo2021}. The multiple activation is not detected in quasistationary simulations of the SIRS dynamics with $\alpha/\mu\le 1$. Top panels of Fig.~\ref{fig:hub+RRN} present the estimated epidemic thresholds as functions of the network size for three values of $\alpha/\mu$ considering QS simulation, PQMF and rDMP mean-field theories. The corresponding thresholds for a pure RR network are also presented. Since the hub size is fixed, epidemic thresholds of QS simulations converge to the value obtained in  the pure RR network in the thermodynamic limit. However, localization remains relevant at finite-size systems altering the convergence to the asymptotic limit: While the pure RR presents a monotonic decay towards the asymptotic value, the presence of the hub lowers the threshold and inverts the finite-size dependence. In contrast with the pure RR networks shown in Fig.~\ref{fig:RRN}, the PQMF theory deviates significantly from  the simulation outcomes, the more for higher rate of waning immunity, being thus outperformed by rDMP. 

The localization associated with the mean-field theories and simulations characterized with the IPR of the Jacobian‘s PEV and NAV at the threshold, respectively,  are shown in Figs.~\ref{fig:hub+RRN}(d,e,f). The PQMF theory presents a finite IPR due to the localization in the hub, while the NAV obtained in simulation becomes delocalized as the network size increases. Notice, however, that the IPR decays slower with size than predicted by the Jacobian of PEV (Hashimoto matrix) in the rDMP theory, being differences more evident for larger $\alpha$, showing that the actual stochastic dynamics is more localized than that of the rDMP.  Indeed, the PEV of the Hashimoto matrix for the RR network with an integrated hub is localized only  if $K\gg(N/\av{n})^{1/2}$~\cite{Pastor_Satorras2020}.

\subsection{Power-law networks}

\begin{figure}[hbt!]
	\includegraphics[width=\linewidth]{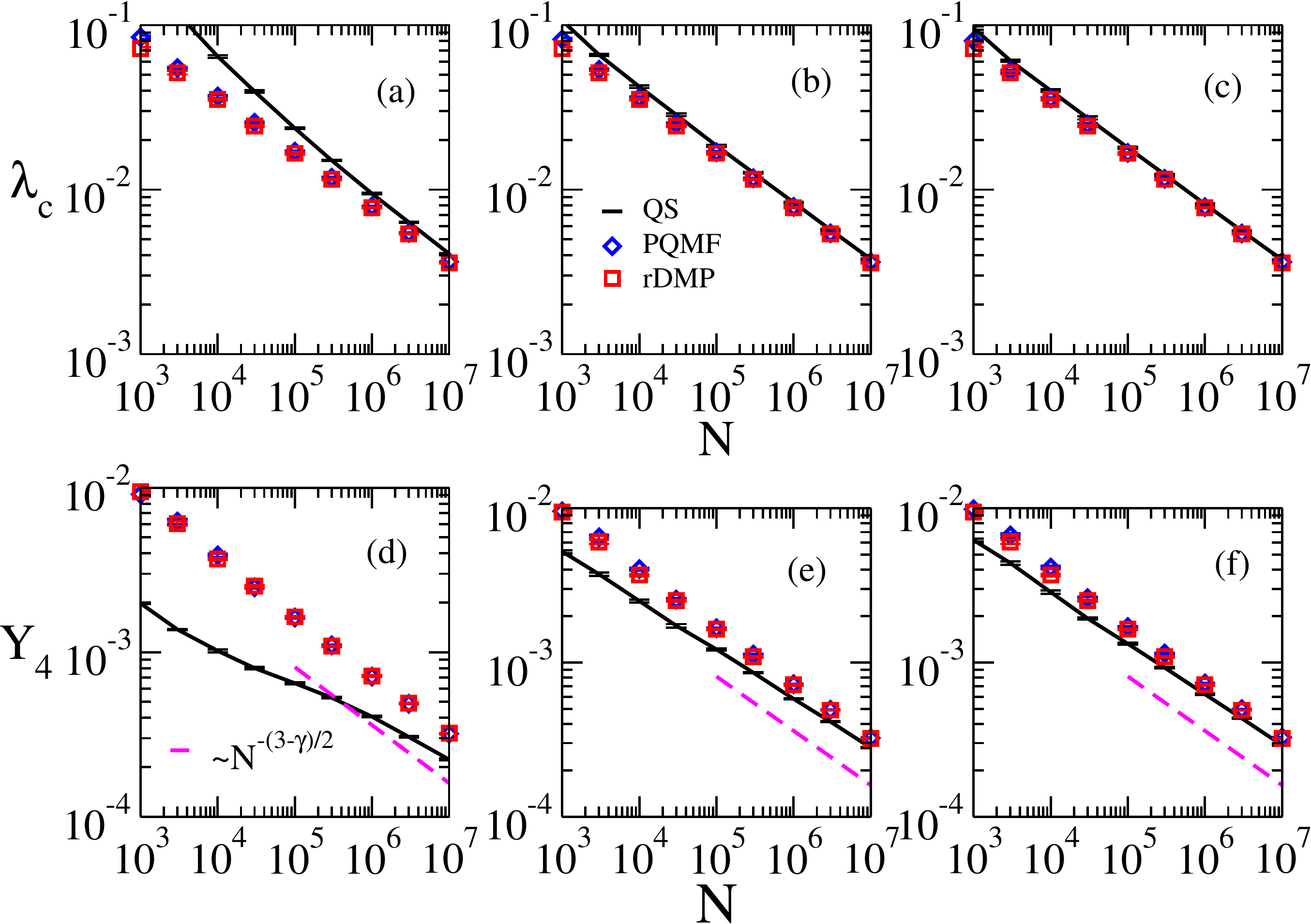}
	\caption{(a,b,c) Epidemic threshold  and (d,e,f) IPR  as functions of the network size for the SIRS dynamics on UCM networks with $\gamma=2.3$ using (a,d)$~\alpha/\mu=0.1$, (b,e)$~0.5$ and (c,f)$~1.0$. Stochastic simulations (black line) are compared to PQMF (blue diamonds) and rDMP (red squares). Dashed line is a guide to eyes indicating the scaling $Y_{4}\sim N^{(3-\gamma)/2}$, expected for IPR of vector localized in maximum K-core~\cite{Satorras2016}. }
	\label{fig:PL_g230}	
\end{figure}

We investigated SIRS dynamics on synthetic uncorrelated networks presenting a power-law degree distribution, $P(k)\sim k^{-\gamma}$, generated through the UCM model~\cite{Catanzaro2005} with a structural cutoff  $k_{\text{c}}=2\sqrt{N}$. The threshold and IPR analyzes for power-law networks with $\gamma<5/2$ are presented in Fig.~\ref{fig:PL_g230}. For all investigated values of $\alpha$, the same behavior is observed: the epidemic threshold goes to zero in simulations as well as PQMF and rDMP theories, the last two being indistinguishable from each other in the presented scales. Simulations asymptotically agree with mean-field theories being the convergence faster for higher waning of immunity. The localization analyses indicate the agreement between simulations and mean-field theories, whose IPR scales consistently with an epidemic localization in the  maximum K-core, as conjectured for SIRS dynamics in this range of degree exponent $\gamma$~\cite{Ferreira2016}. The maximum K-core is a strongly connected subgraph obtained by means of a K-core decomposition~\cite{Dorogovtsev2006}. Thus, our results for SIRS support that the outbreak is triggered as does the SIS dynamics for $\gamma<2/5$~\cite{Castellano2012}. 

For $\gamma>5/2$, the PEV of the PQMF’s Jacobian is localized in the largest hub and its neighbors~\cite{Diogo2019}, as does the adjacency matrix~\cite{Satorras2016}, differently from the Hashimoto matrix whose PEV $v_{i} \sim \sum_{j}A_{ij}(k_{j}-1)$~\cite{Pastor_Satorras2020} leads to a different type of localization. In both cases the respective LEVs diverge for $5/2<\gamma<3$, but following different scaling laws. When $\gamma>3$, the LEV of the PQMF’s Jacobian still diverges in the thermodynamic limit~\cite{Mata2013} and remains finite for the Hashimoto matrix~\cite{Pastor_Satorras2020}. Thus,  rDMP and PQMF  theories predict, respectively, finite and null epidemic thresholds for $\gamma>3$. For this reason, we analyze the case $\gamma=3.5$ where differences are more noticeable. 

\begin{figure}[hbt!]
	\includegraphics[width=\linewidth]{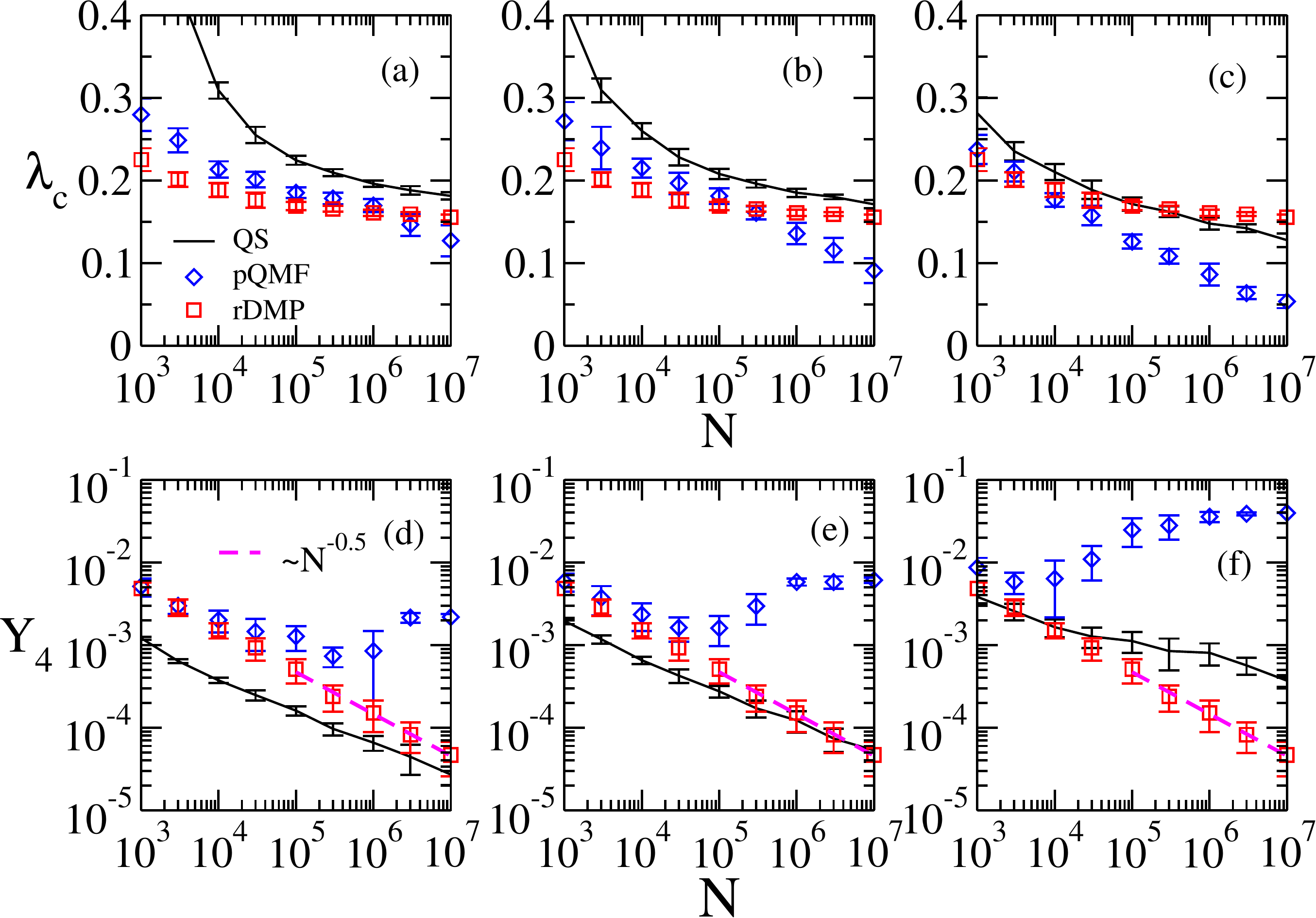}
	\caption{(a,b,c) Epidemic threshold  and (d,e,f) IPR as functions of the network size for a SIRS dynamics on UCM networks with $\gamma=3.5$ using (a,d)$~\alpha=0.1$, (b,e)$~0.5$, and (c,f)$~1.0$. Stochastic simulations (black line) are compared to PQMF (blue diamonds) and rDMP (red squares).}
	\label{fig:PL}
\end{figure}

The finite-size scaling of the epidemic threshold of stochastic simulations depends on the rate of waning of immunity while the asymptotic threshold decreases only slightly  with $\alpha$,  as shown at the top panels of Fig.~\ref{fig:PL}.  In the range of network sizes investigated (up to $N=10^{7}$), the epidemic threshold seems to converge to a finite value, which is qualitatively described by rDMP theory. The PEV associated to the PQMF’s Jacobian matrix is strongly localized in some  nodes represented by an asymptotically finite IPR, the stronger for larger $\alpha$. Conversely, the PEV of the Hashimoto matrix does not depend on $\alpha$, being localized in a subextensive fraction of nodes manifested as a scaling law $Y_{4}\sim N^{-a}$, with $a<1$. Stochastic simulations present a  localization pattern which depends on $\alpha$, becoming slightly more localized as the immunity time $1/\alpha$ decreases. The PQMF is clearly outperformed by the rDMP theory. However, simulations indicate that rDMP theory yields an epidemic activity less localized than the actual simulations and overestimate the asymptotic epidemic threshold, more evident for larger $\alpha$.

\section{Conclusions}
\label{sec:conclusions}

The development of theoretical frameworks capable of reproducing with accuracy epidemic models is crucial to the progress of forecasting and controlling epidemic outbreaks. Basic models such as SIS and SIR present different natures of epidemic activation and are better suited into different theoretical approaches. Waning immunity with the rate $\alpha$, where a recovered individual becomes susceptible again after an average time $1/\alpha$, introduces the generalized SIRS  dynamics, which interpolates between  SIR  $(\alpha\rightarrow 0)$ and SIS $(\alpha\rightarrow\infty)$ epidemic models. While some aspects of SIRS dynamics are akin to the SIS model (existence of a active steady state and universality class in regular lattices~\cite{DeSouza2010,Joo2004}) others resemble the SIR dynamics (the finite epidemic threshold for degree exponent $\gamma>3$ and activation mechanism in complex networks~\cite{Ferreira2016}). So, the mean-field theory that better describes the SIRS dynamics is not completely ascertained. The present work investigates the SIRS model within two theoretical frameworks, namely rDMP or PQMF theories, and stochastic simulations. Both theories are pairwise approaches, while in rDMP does not permit backtracking reinfection, in which an infected node can infect the neighbor that infected itself, while PQMF does. We tackle the problem of which mean-field theory more accurately reproduces the epidemic threshold and epidemic localization patterns of the SIRS dynamics on different types of complex networks.

In the case of the homogeneous degree distribution, where no relevant localization is present, PQMF theory outperforms rDMP. However, the introduction of an immersed single node of a large, but size-independent degree, promotes strong localization effects and the PQMF performance becomes worse. The rDMP for SIRS dynamics on star graphs, which play the role of isolated hubs immersed into a network, indeed predict a finite epidemic lifespan independently of the star size, while an exponential divergence with size is obtained for the PQMF theory. Neither rDMP or PQMF theories predict the algebraic increase with the star size reported in Ref.~\cite{Ferreira2016}, indicating that these theories under or overestimate, respectively, the localization of the epidemic activity around hubs in the networks. The SIRS dynamics on networks with power-law degree distribution confirms this conjecture. Indeed, a finite epidemic threshold observed in simulations of  the SIRS dynamics on networks with degree exponent $\gamma>3$ is in qualitative agreement with rDMP theory and contrasts with the vanishing epidemic threshold obtained with the PQMF theory. However, the localization analysis also points out that rDMP underestimates the actual epidemic localization observed in simulations, occurring in a subextensive fraction of the network that is asymptotically much larger than the subset corresponding to the  localization of the PEV of the Hashimoto matrix predicted by the rDMP theory. 

Our results call for modified versions of the rDMP theory, which softens the strict prohibition of backtracking reinfection to predict more accurately the localization pattern and thus the epidemic threshold of SIRS dynamics in networks. We also expect the results presented here to be applied to more complex dynamical models on networks.
 
\appendix

\section{Stochastic simulation of the SIRS model}
\label{app:algoritmo}

We performed stochastic simulations of the SIRS model using an optimized Gillespie algorithm~\cite{Cota2017}. Lets defined the number of recovered $N_{\text{rec}}$ and infected $N_{\text{inf}}$ nodes as well as  the total number of edges emanating from the latter, $N_\text{SI}$. At each time step, with probability       
\begin{equation}
	P_{\text{I$\rightarrow$R}}=\frac{\mu \Ninf}{\mu\Ninf+\lambda\NSI+\alpha N_{\text{rec}}},
	\label{eq:GA1}
\end{equation}
an infected node is chosen at random and recovered. With probability 
\begin{equation}
P_{\text{R$\rightarrow$S}}=\frac{\alpha N_{\text{rec}}}{\mu\Ninf+\lambda\NSI+\alpha N_{\text{rec}}},
\label{eq:GA2}
\end{equation}
a recovered node is chosen at random and becomes  susceptible. Finally, with probability
\begin{equation}
	P_\text{I$\rightarrow$S}=\frac{\lambda\NSI}{\mu\Ninf+\lambda\NSI+\alpha N_{\text{rec}}},
	\label{eq:GA3}
\end{equation}
an infected node $i$ is selected with probability proportional to its degree $k_i$. Then, a neighbor of $i$ is chosen at random and becomes infected if it is susceptible; otherwise the simulation goes to next step without changing the configuration. Finally, time is incremented by 
\begin{equation}
\delta t  = \frac{-\ln u}{\mu\Ninf+\lambda\NSI},
\end{equation}
where $u$ is a pseudo random number uniformly distributed in the interval $(0,1)$.

A finite system always falls into the absorbing state if the simulation runs for a time long enough~\cite{Oliveira2005}. This feature can be handled using a scheme known as standard quasistationary method~\cite{Oliveira2005,Sander2016,Costa2021}. A list of $M$ configurations is built and constantly updated replacing one of its configurations, selected at random, by the current one  with probability $P_{\text{rep}}$ by unit time. We used $M=50$ and  $P_{\text{rep}}=0.01$ in the present work. The quasi-stationary averages were computed over a time window varying from $t_\text{av}=10^5$ to $2\times10^6$, after a relaxation time of  $t_\text{rlx}=10^5$ time units. The longest time intervals were used for the lowest densities, where fluctuations are more relevant.

\section{QMF theory for the SIRS dynamics}
\label{app:critical_qmf}

To determine the  critical quantities in teh QMF theory we perform a linear stability analysis around the absorbing state $\rho_i=0$. In the steady-state, Equation~\eqref{eq:PQMF2b} leads to
\begin{eqnarray}
s_{i}=1-\left(1+\frac{\mu}{\alpha}\right)\rho_{i} \label{eq:AQMF1}.
\end{eqnarray}	
We perform a quasi-static approximation, plugging this result into Eq.~\eqref{eq:QMFa} to obtain
\begin{eqnarray}
\frac{d\rho_{i}}{dt}\approx-\mu \rho_{i}+\lambda \left[1-\left(1+\frac{\mu}{\alpha}\right)\rho_{i}\right]\sum_{j} A_{ij}\rho_{j},
\label{eq:AQMF2}
\end{eqnarray}
which, in leading order in $\rho_i$, becomes
\begin{eqnarray}
\frac{d\rho_{i}}{dt}=\sum_{j}L_{ij}\rho_{j},
\label{eq:AQMF3a}	
\end{eqnarray}
where $L_{ij}$ is the Jacobian matrix given by 
\begin{eqnarray}
L_{ij}=-\mu \delta_{ij}+\lambda\sum_{j} A_{ij}.		
\label{eq:AQMF3b}
\end{eqnarray}
The absorbing state losses stability when the largest eigenvalue of the Jacobian is null and, therefore, the threshold of SIRS model assumes the form  
\begin{equation}
\lambda_{c}^{\text{SIRS}}=\frac{\mu}{\Lambda^{(1)}}.
\label{eq:AQMF4}
\end{equation}
in which, $\Lambda^{(1)}$ is the LEV of the adjacency matrix. The steady-state of Eq.~\eqref{eq:AQMF2} yields
\begin{equation}
\rho_{i}=\frac{\lambda \sum_{j}A_{ij} \rho_{j}}{\mu +\lambda\left(1+\frac{\mu}{\alpha}\right)\sum_{j}A_{ij}\rho_{j}}. 
\label{eq:AQMF5}
\end{equation} 
The epidemic prevalence $\rho_i$ can be expanded in terms of eigenvectors $\{v_i^{(l)}\}$ of $A_{ij}$~\cite{Goltsev2012}, $\sum_j A_{ij}v_j^{(l)} = \Lambda^{(l)} v_i^{(l)}$, where  $v_i^{(1)}$ corresponds to the PEV, $v_i^{(2)}$ to the eigenvector with second LEV, and so on. Assuming a spectral gap $\Lambda^{(1)}\gg\Lambda^{(l)}$, $l>1$, near the epidemic threshold where $\rho_i\ll 1$, we obtain 
\begin{equation}
	\rho_{i} \approx c^{(1)}v^{(1)}_{i}.
	\label{eq:rhoiapprox}
\end{equation}
to the leading order in $\rho_i$, Plugging Eqs.~\eqref{eq:rhoiapprox} and \eqref{eq:AQMF5} leads to
\begin{eqnarray}
\frac{\lambda}{\mu} \Lambda^{(1)}\sum_{i}^{N}\frac{\left[v_{i}^{(1)}\right]^{2}}{1+\frac{\lambda}{\mu}\left(1+\frac{\mu}{\alpha}\right)\Lambda^{(1)}c^{(1)}v_{i}^{(1)}}\simeq 1.
\label{eq:AQMF6}	
\end{eqnarray}
Expanding Eq.~\eqref{eq:AQMF6} for $\rho_i\approx c^{(1)}v_i^{(1)}\ll 1$,noting that $\frac{\lambda}{\mu} \Lambda^{(1)}$ is $\mathcal{O}(1)$, we obtain
\begin{equation}
c^{(1)} \simeq \frac{\frac{\lambda}{\mu}\Lambda^{(1)}-1}{\left(1+\frac{\mu}{\alpha}\right)\sum_i\left[v_i^{(1)}\right]^3},
\label{eq:c1}
\end{equation}
which is used to compute the epidemic prevalence as
\begin{equation}
\rho =  \frac{1}{N} \sum_{i} \rho_i \simeq \frac{\alpha}{\mu+\alpha}  a(N)
 \frac{\lambda\Lambda^{(1)}-\mu}{\mu}
 \label{eq:rhoqmf}
\end{equation}
where the pre-factor $a(N)$ is a function of $N$ given by
\begin{equation}
a(N) = \frac{\sum_i v_i^{(1)}}{N\sum_i\left[v_i^{(1)}\right]^3},
\end{equation}
implying that $\rho \sim (\lambda-\lambda_\text{c})^\beta$ with critical exponent $\beta=1$. Comparing Eq.~\eqref{eq:rhoqmf} with the QMF solution of the SIS model presented in Ref.~\cite{Goltsev2012}, we obtain a proportionality relation between SIRS and SIS prevalences given by
\begin{equation}
\rho^{\text{SIRS}}\simeq\left(\frac{\alpha}{\mu+\alpha}\right)\rho^{\text{SIS}},
\end{equation}
implying that QMF theory predicts the same critical properties for SIRS and SIS models.

\begin{acknowledgments}

DHS thanks the support given by \textit{Fundação de Amparo à Pesquisa do Estado de São Paulo} (FAPESP)-Brazil (Grants no. 2021/00369-0 and 2013/07375-0).
SCF thanks the support by the \textit{Conselho Nacional de Desenvolvimento Científico e Tecnológico} (CNPq)-Brazil (Grants no. 430768/2018-4 and 311183/2019-0) and \textit{Fundação de Amparo à Pesquisa do Estado de Minas Gerais} (FAPEMIG)-Brazil (Grant no. APQ-02393-18).  FAR acknowledges CNPq (grant 309266/2019- 0) and FAPESP (Grant 19/23293-0) for the financial support given for this research.  This study was financed in part by the \textit{Coordena\c{c}\~{a}o de Aperfei\c{c}oamento de Pessoal de N\'{i}vel Superior} (CAPES) - Brazil - Finance Code 001.
\end{acknowledgments}

%
 
\end{document}